\documentclass[epj,twocolumn]{webofc}
\usepackage[varg]{txfonts}   
\woctitle{Dark Matter, Hadron Physics and Fusion Physics}
\begin{document}
\boldmath
\title{Search for the dark photon in $\pi^0$ decays by the NA48/2 experiment at CERN}
\unboldmath

\author{Evgueni Goudzovski\inst{1}\fnsep\thanks{For the NA48/2 Collaboration: G.~Anzivino, R.~Arcidiacono, W.~Baldini, S.~Balev, J.R.~Batley, M.~Behler,
S.~Bifani, C.~Biino, A.~Bizzeti, B.~Bloch-Devaux
% ...
}}

\institute{School of Physics and Astronomy, University of Birmingham, United Kingdom}

\abstract{%
A sample of $4.687\times 10^6$ fully reconstructed $K^\pm\to\pi^\pm\pi^0_D$, $\pi^0_D\to\gamma e^+e^-$ decay candidates in the kinematic range $m_{ee}>10~{\rm MeV}/c^2$ with a negligible background contamination collected by the NA48/2 experiment at CERN in 2003--04 is analysed to search for the dark photon ($A'$) via the decay chain $K^\pm\to\pi^\pm\pi^0$, $\pi^0\to\gamma A'$, $A'\to e^+e^-$. No signal is observed, and preliminary exclusion limits on space of dark photon mass $m_A'$ and mixing parameter $\varepsilon^2$ are reported.
}
\maketitle
\section{Introduction}
\label{sec:intro}

Kaons represent a source of tagged neutral pion decays, mainly via their $K^\pm\to\pi^\pm\pi^0$ and $K_L\to3\pi^0$ decays. Therefore high intensity kaon experiments provide opportunities for precision studies of $\pi^0$ decay physics. One of them is the NA48/2 experiment at the CERN SPS, which collected a large sample of charged kaon ($K^\pm$) decays in flight in 2003--04 corresponding to about $2\times 10^{11}$ $K^\pm$ decays in the vacuum decay volume~\cite{ba07}.

The large sample of $\pi^0$ mesons produced and decaying in vacuum collected by NA48/2 allows for a high sensitivity search for the dark photon ($A'$), a hypothetical gauge boson appearing in hidden sector new physics models with an extra $U(1)$ gauge symmetry. In a rather general set of models, the interaction of the DP with the visible sector is through kinetic mixing with the Standard Model hypercharge $U(1)$~\cite{ho86}. In these models, the new coupling constant $\varepsilon$ is proportional to the electric charge and the DP couples in exactly the same way to quarks and leptons. These scenarios provide an explanation to the observed rise in the cosmic-ray positron fraction with energy, and offer an explanation to the muon gyromagnetic ratio ($g-2$) problem~\cite{po09}.

From the experimental point of view, the DP is characterized by two a priori unknown parameters, the mass $m_{A'}$ and the mixing parameter $\varepsilon$. Its possible production in the $\pi^0$ decay and subsequent decay proceed via the following chain:
\begin{displaymath}
K^\pm\to\pi^\pm\pi^0, ~~~ \pi^0\to\gamma A', ~~~ A'\to e^+e^-,
\end{displaymath}
with three charged particles and a photon in the final state. The expected branching fraction of the $\pi^0$ decay is~\cite{batell09}
\begin{equation}
{\cal B}(\pi^0\to\gamma A') = 2\varepsilon^2 \left(1-\frac{m_{A'}^2}{m_{\pi^0}^2}\right)^3 {\cal B}(\pi^0\to\gamma\gamma),
\label{eq:br}
\end{equation}
with a kinematic suppression of the decay rate for high DP masses $m_{A'}$ approaching $m_{\pi^0}$. In the mass range $2m_e\ll m_{A'}<m_{\pi^0}$ accessible in this analysis, the DP is below threshold for all decays into SM fermions except $A'\to e^+e^-$, while the allowed loop-induced decays ($A'\to 3\gamma$, $A'\to\nu\bar\nu$) are highly suppressed. Therefore, assuming that the DP decays only into SM particles, ${\cal B}(A'\to e^+e^-)\approx 1$. The expected total DP decay width is then~\cite{batell09-brs}
\begin{displaymath}
\Gamma(A'\to e^+e^-) = \frac{1}{3}\alpha\varepsilon^2 m_{A'} \sqrt{1-\frac{4m_e^2}{m_{A'}^2}}\left(1+\frac{2m_e^2}{m_{A'}^2}\right).
\end{displaymath}
It follows that the DP mean proper lifetime for $2m_e\ll m_{A'}<m_{\pi^0}$ is
\begin{displaymath}
c\tau \approx 0.8~{\mu\rm m} \times \left(\frac{10^{-6}}{\varepsilon^2}\right) \times \left(\frac{100~{\rm MeV}}{m_{A'}}\right).
\end{displaymath}
The maximum DP mean path in the NA48/2 reference frame in a fully reconstructed event corresponds to an energy of approximately $E_{\rm max}=50~{\rm GeV}$:
\begin{displaymath}
L_{\rm max} \approx (E_{\rm max}/m_{A'}) c\tau \approx 0.4~{\rm mm} \times \left(\frac{10^{-6}}{\varepsilon^2}\right) \times \left(\frac{100~{\rm MeV}}{m_{A'}}\right)^2,
\end{displaymath}
It does not exceed 10~cm and can be neglected for $m_{A'}>10$~MeV/$c^2$ and $\varepsilon^2>5\times 10^{-7}$. The DP in the above parameter range can be assumed to decay at the production point ({\it prompt decay}), and the 3-track vertex topology can used without significant acceptance loss. The DP signature is identical to that of the Dalitz decay $\pi^0_D\to e^+e^-\gamma$, which therefore represents an irreducible background and determines the sensitivity. The largest $\pi^0_D$ sample, and therefore the largest sensitivity, is obtained from the $K^\pm\to\pi^\pm\pi^0_D$ decays (denoted $K_{2\pi D}$ below).

\begin{figure}[t]
\begin{center}
\resizebox{0.45\textwidth}{!}{\includegraphics{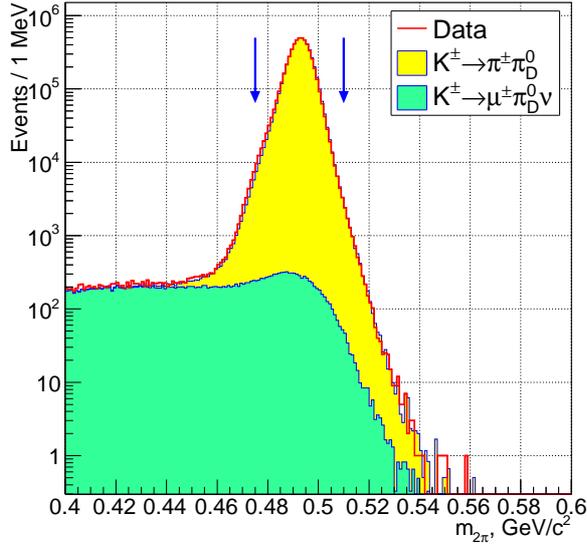}}%
\end{center}
\vspace{-7mm}
\caption{Reconstructed $\pi^\pm\pi^0_D$ invariant mass ($m_{2\pi}$) distributions of the data and simulated $K_{2\pi D}$ and $K_{\mu 3D}$ samples. The selection condition $475~{\rm MeV}/c^2<m_{2\pi}<510~{\rm MeV}/c^2$ is illustrated with arrows.}
\label{fig:mass}
\end{figure}

\section{The NA48/2 experiment}
\label{sec:experiment}

The NA48/2 beam line has been designed to deliver simultaneous narrow momentum band $K^+$ and $K^-$ beams following a common beam axis derived from the primary 400 GeV/$c$ protons extracted from the CERN SPS. Secondary beams with central momenta of $(60\pm3)~{\rm GeV}/c$ (r.m.s.) were used. The beam kaons decayed in a fiducial decay volume contained in a 114~m long cylindrical vacuum tank. The momenta of charged decay products were measured in a magnetic spectrometer, housed in a tank filled with helium placed after the decay volume. The spectrometer comprised four drift chambers (DCHs), two upstream and two downstream of a dipole magnet which provided a horizontal transverse momentum kick of $120~\mathrm{MeV}/c$ to singly-charged particles. Each DCH was composed of eight planes of sense wires. A plastic scintillator hodoscope (HOD) producing fast trigger signals and providing precise time measurements of charged particles was placed after the spectrometer. Further downstream was a liquid krypton electromagnetic calorimeter (LKr), an almost homogeneous ionization chamber with an active volume of 7~m$^3$ of liquid krypton, $27X_0$ deep, segmented transversally into 13248 projective $\sim\!2\!\times\!2$~cm$^2$ cells and with no longitudinal segmentation. The LKr information is used for photon measurements and charged particle identification. An iron/scintillator hadronic calorimeter and muon detectors were located further downstream. A dedicated two-level trigger was in operation to collect three-track decays (including $K_{2\pi D}$ used for this study) with an efficiency of about 98\%. A detailed description of the detector can be found in Ref.~\cite{fa07}.

\section{Event selection}
\label{sec:selection}

The full NA48/2 data sample is used for the analysis. The $K_{2\pi D}$ event selection requires a three-track vertex reconstructed in the fiducial decay region formed of a pion ($\pi^\pm$) candidate track and two opposite-sign electron ($e^\pm$) candidate tracks. Charged particle identification is based on the ratio of energy deposition in the LKr calorimeter to the momentum measured by the spectrometer, which should be smaller (greater) than 0.85 for pion (electron) candidates. Furthermore, a single insolated LKr energy deposition cluster is required and considered as the photon candidate. A set of tight selection criteria is applied to the energies of the final state particles, their timing and spatial separations. The total reconstructed $\pi^\pm\pi^0$ momentum is required to be consistent with the beam momentum spectrum, and its transverse component with respect to the nominal beam axis is required to be consistent with no missing momentum. The reconstructed invariant mass of the $\pi^\pm\pi^0$ system (Fig.~\ref{fig:mass}) is required to the consistent with the $K^\pm$ mass.

\begin{figure}[t]
\begin{center}
\resizebox{0.45\textwidth}{!}{\includegraphics{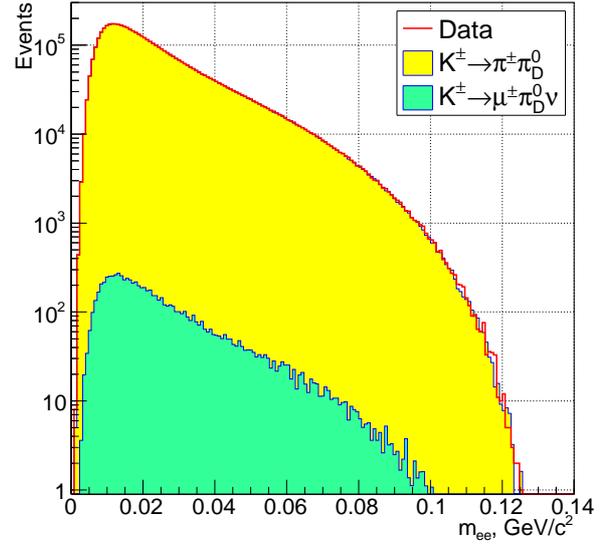}}
\end{center}
\vspace{-7mm}
\caption{Reconstructed $e^+e^-$ invariant mass distributions of the data and simulated $K_{2\pi D}$ and $K_{\mu 3D}$ samples.}
\label{fig:mee}
\end{figure}

A sample of $4.687\times 10^6$ fully reconstructed $\pi^0_D$ decay candidates in the $e^+e^-$ invariant mass range $m_{ee}>10~{\rm MeV}/c^2$ with a negligible background is selected. The candidates mainly originate from $K_{2\pi D}$ decays, with 0.15\% coming from the semileptonic $K^\pm\to\pi^0_D\mu^\pm\nu$ decays (denoted $K_{\mu 3 D}$ below). Correcting the observed number of candidates for acceptance and trigger efficiency, the total number of $K^\pm$ decays in the 98~m long fiducial decay region for the analysed data sample is found to be $N_K=(1.55\pm0.05)\times 10^{11}$, where the quoted error is dominated by the external uncertainty on the $\pi^0_D$ decay branching fraction ${\cal B}(\pi^0_D)$. The reconstructed $e^+e^-$ invariant mass ($m_{ee}$) spectrum of the $K_{2\pi D}$ candidates is displayed in Fig.~\ref{fig:mee}. A dark photon produced in the $\pi^0_D$ decay and decaying promptly to $e^+e^-$ would be produce a narrow spike in the spectrum.

\section{Background simulation}
\label{sec:background}

Monte Carlo (MC) simulations of the $K_{2\pi D}$ and $K_{\mu 3D}$ processes are performed to subtract the irreducible $\pi^0_D$ background. The $\pi^0_D$ decay is simulated using the lowest-order differential decay rate~\cite{mi72}
\begin{displaymath}
\frac{d^2\Gamma}{dxdy} = \Gamma_0\frac{\alpha}{\pi}|F(x)|^2\frac{(1-x)^3}{4x}\left(1+y^2+\frac{r^2}{x}\right),
\end{displaymath}
where $r = 2m_e/m_{\pi^0}$, the kinematic variables are
\begin{displaymath}
x = \frac{(q_1+q_2)^2}{m_\pi^2} = (m_{ee}/m_{\pi^0})^2, ~~~~~ y = \frac{2p(q_1-q_2)}{m_{\pi^0}(1-x)},
\end{displaymath}
$q_1$, $q_2$ and $p$ are the four-momenta of the electrons ($e^\pm$) and the pion ($\pi^0$), $\Gamma_0$ is the rate of the $\pi^0\to\gamma\gamma$ decay, and $F(x)$ is the pion transition form factor (TFF). The differential decay rate falls steeply as a function of $x$.

Radiative corrections to the differential rate are implemented following the classical approach of Mikaelian and Smith~\cite{mi72} revised recently to improve the numerical precision~\cite{husek}: the differential decay rate is modified using
\begin{displaymath}
\frac{d^2\Gamma^{\rm rad}}{dxdy} = \delta(x,y) \cdot \frac{d^2\Gamma}{dxdy},
\end{displaymath}
which does not account for the emission of inner bremsstrahlung photons.

The TFF is conventionally parameterized as $F(x)=1+ax$. The TFF slope parameter $a$ is expected from vector meson dominance models to be $a\approx 0.03$, and detailed theoretical calculations based on dispersion theory yield $a = 0.0307\pm0.0006$~\cite{ho14}. Experimentally, the PDG average $a=0.032\pm0.004$~\cite{pdg} is determined mainly from a $e^+e^-\to e^+e^-\pi^0$ rate measurement in the space-like region by the CELLO experiment~\cite{cello}, while the measurements from the $\pi^0$ decay have very limited accuracy. The precision on the used radiative corrections to the $\pi^0_D$ decay is limited: in particular, the missing correction to the measured TFF slope due to two-photon exchange is estimated to be $\Delta a=+0.005$~\cite{ka06}. Therefore the background description cannot benefit from the precise inputs on the TFF slope~\cite{ho14, pdg}, and an ``effective'' TFF slope obtained from a fit to the data $m_{ee}$ spectrum itself is used to obtain a satisfactory background description (as quantified by a $\chi^2$ test) over the range $m_{ee}>10~{\rm MeV}/c^2$ used for the DP search. The low $m_{ee}$ region is not considered for the search as the acceptance computation is less robust due to the steeply falling geometrical acceptance at low $m_{ee}$ and decreasing electron identification efficiency at low momentum.

\begin{figure}[t]
\begin{flushright}
\vspace{-5mm}
\resizebox{0.45\textwidth}{!}{\includegraphics{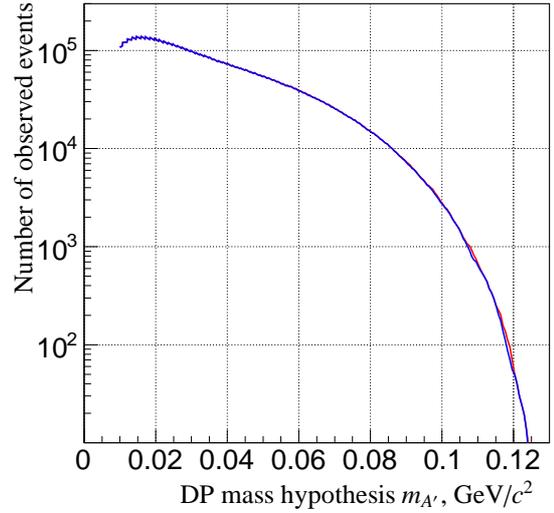}}%
\put(-160, -2){DP mass hypothesis $m_{A'}$, GeV/$c^2$}
\put(-222, 77){\rotatebox{90}{Number of observed events}}
\end{flushright}
\vspace{-3mm}
\caption{Numbers of observed (red) and expected (blue) DP candidates in the DP signal mass window as for each assumed DP mass hypothesis, which are used as input values for the Rolke--L\'opez confidence interval calculation. The two curves are hardly distinguishable in the logarithmic scale.}
\label{fig:observed}
\end{figure}


\section{Dark photon search}
\label{sec:dp}

A DP mass scan (i.e. a search for the DP assuming different mass hypotheses with a variable mass step) is performed. The mass step of the scan and the width of the DP signal mass window around the assumed mass are determined by the resolution $\delta m_{ee}$ on the $e^+e^-$ invariant mass $m_{ee}$. The mass-dependent resolution as a function of $m_{ee}$ evaluated from a $K_{2\pi D}$ MC sample is parameterized as $\sigma_{m} = 0.067~{\rm MeV}/c^2 + 0.0105\cdot m_{ee}$. The mass step of the DP scan is set to be $\sigma_{m}/2$, while the signal region mass window for each DP mass hypothesis is defined as $\pm1.5\sigma_m$ around the assumed mass (both the scan step and the mass window half-width are rounded to the nearest multiple of 0.02 MeV/$c^2$). The mass window width has been optimised with MC simulations to obtain the highest sensitivity to the DP signal, determined by a trade-off between $\pi^0_D$ background fluctuation and signal acceptance.

In total, 398 DP mass hypotheses are tested in the range $10~{\rm MeV}/c^2 \le m_{ee} < 125~{\rm MeV}/c^2$. The lower extent of the considered mass range is determined by the limited precision of MC simulation of background at low mass, while at the upper limit of the mass range the signal acceptance drops to zero. The numbers of observed data events in the signal region ($N_{\rm obs}$) and the numbers of $\pi^0_D$ background events expected from MC simulation corrected by the measured trigger efficiencies ($N_{\rm exp}$) in the DP signal window for each considered mass hypothesis are presented in Fig.~\ref{fig:observed}. They decrease with the DP mass due to the steeply falling $\pi^0_D$ differential decay rate and decreasing acceptance, even though the mass window width increases, being approximately proportional to the mass.

\begin{figure}[t]
\vspace{-5mm}
\begin{flushright}
\resizebox{0.47\textwidth}{!}{\includegraphics{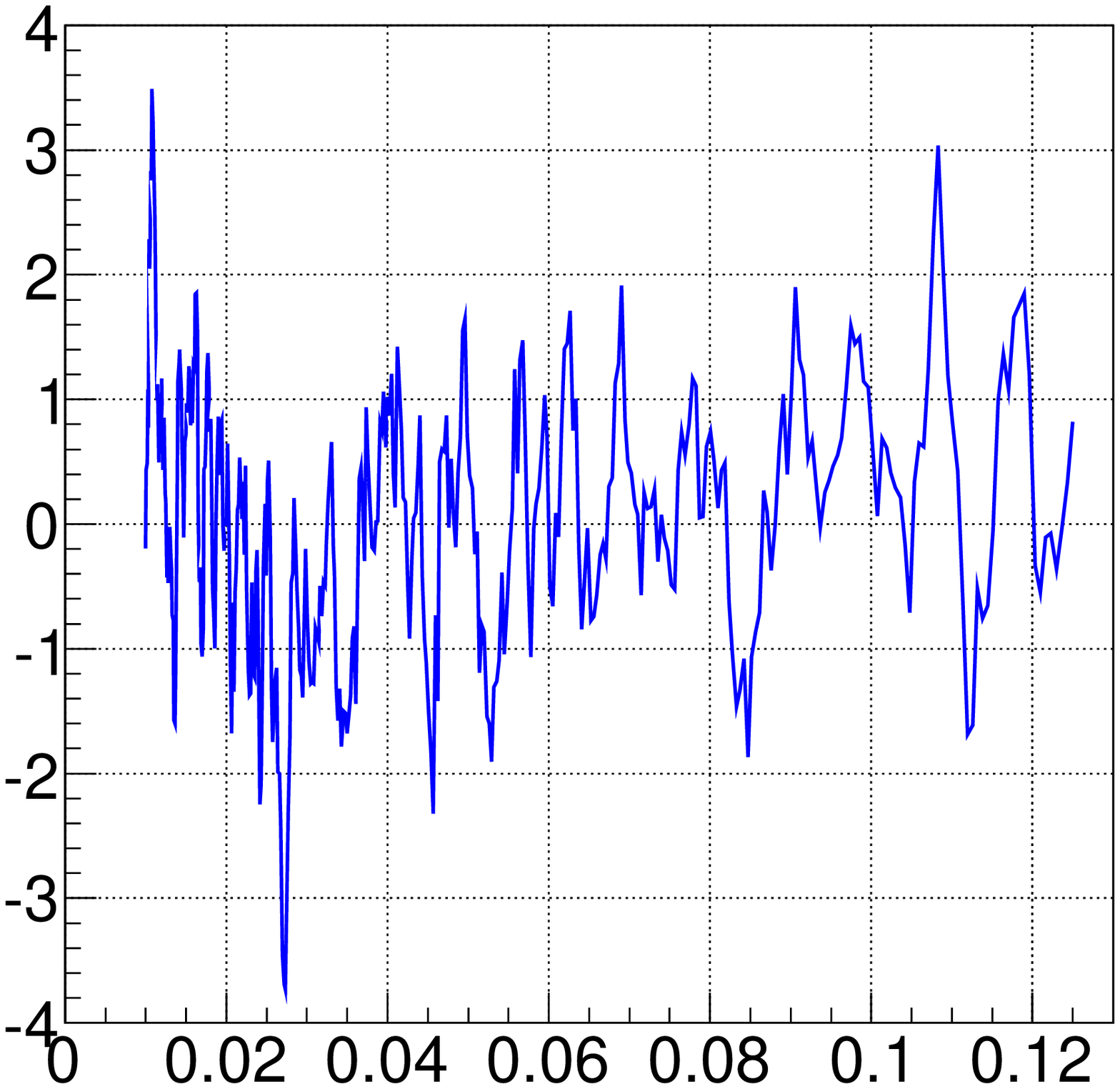}}%
\put(-160, -1){DP mass hypothesis $m_{A'}$, GeV/$c^2$}
\put(-225, 70){\rotatebox{90}{Dark photon signal significance}}
\end{flushright}
\vspace{-5mm}
\caption{Statistical significance of the DP signal for each mass hypothesis.}
\label{fig:sig}
\end{figure}

The statistical significance of the DP signal in each mass hypothesis is defined as
\begin{displaymath}
S=(N_{\rm obs}-N_{\rm exp})/\sqrt{(\delta N_{\rm obs})^2+(\delta N_{\rm exp})^2},
\end{displaymath}
where $\delta N_{\rm obs}=\sqrt{N_{\rm obs}}$ is the statistical uncertainty on the number of observed events, and $\delta N_{\rm exp}$ is the (larger) uncertainty on the number of expected background events. The latter comes mainly from the statistical errors on the measured trigger efficiencies in the DP signal mass window, with a smaller contribution due to the limited size of the MC samples used to model the $\pi^0_D$ background. The statistical significances of the DP signal for each considered mass hypothesis are shown in Fig.~\ref{fig:sig}: none of them exceeds 3.5, meaning that no statistically significant DP signal is observed. The results obtained in the neighbouring mass hypotheses are highly correlated, as the signal mass window is about 6 times wider than the mass step of the DP scan.

Confidence intervals at 90\% CL for the number of $A'\to e^+e^-$ decay candidates ($N_{\rm DP}$) in each mass hypothesis ($N_{\rm DP}$) are computed from $N_{\rm obs}$, $N_{\rm exp}$ and $\delta N_{\rm exp}$ using the Rolke--L\'opez method~\cite{rolke} assuming Poissonian (Gaussian) errors on the numbers of observed (expected) events. For the preliminary results, it is assumed conservatively that $N_{\rm obs}=N_{\rm exp}$ in cases when $N_{\rm obs}<N_{\rm exp}$, as the employed implementation of the method (from the ROOT package) has been found to underestimate the upper limits in that case.

Upper limits at 90\% CL on ${\cal B}(\pi^0\to\gamma A')$ in each DP mass hypothesis in the assumption ${\cal B}(A'\to e^+e^-)=1$ (which holds for $m_A'<2m_\mu$ if $A'$ decays to SM fermions only) are computed using the relation
\begin{displaymath}
{\cal B}(\pi^0\to\gamma A') = \frac{N_{\rm DP}}{N_K}
\Big[{\cal B}(K_{2\pi}) A(K_{2\pi}) + {\cal B}(K_{\mu 3}) A(K_{\mu 3}) \Big]^{-1}.
\end{displaymath}
The acceptances $A(K_{2\pi})$ and $A(K_{\mu 3})$ of the employed $K_{2\pi D}$ event selection for the $K_{2\pi}$ and $K_{\mu 3}$ decays, respectively, followed by the prompt $\pi^0\to\gamma A'$, $A'\to e^+e^-$ decay chain, are evaluated for each considered DP mass with MC simulation. Event distributions in the angle between $e^+$ momentum in the $e^+e^-$ rest frame and the $e^+e^-$ momentum in the $\pi^0$ rest frame are identical for the decay chain involving the DP ($\pi^0\to\gamma A'$, $A'\to e^+e^-$) and the $\pi^0_D$ decay, up to a negligible effect of the radiative corrections that should not be applied in the former case (found to influence the acceptance at the level below 1\% in relative terms). Therefore DP acceptances are evaluated using the MC samples produced for background description, and no dedicated MC productions are required.

\begin{figure}[tb]
\begin{flushright}
\resizebox{0.45\textwidth}{!}{\includegraphics{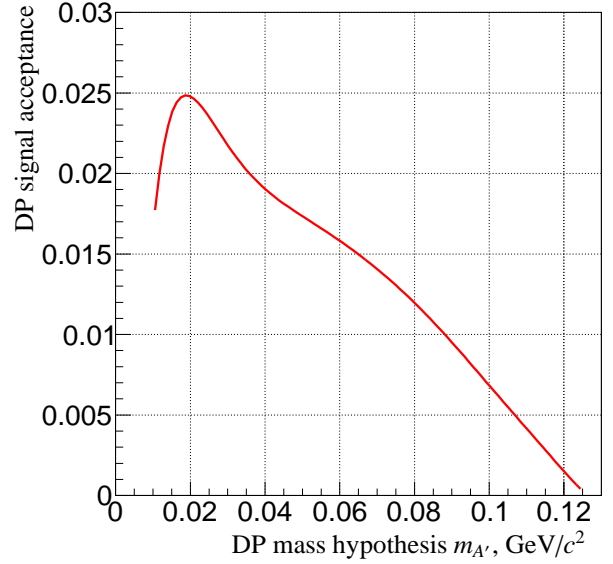}}%
\put(-147, -1){DP mass hypothesis $m_{A'}$, GeV/$c^2$}
\put(-228, 120){\rotatebox{90}{DP signal acceptance}}
\end{flushright}
\vspace{-5mm}
\caption{Acceptance of the event selection for the $K_{2\pi}$, $\pi^0\to\gamma A'$, $A'\to e^+e^-$ decay chain with a prompt $A'$ decay as a function of the assumed DP mass evaluated with MC simulation.}
\label{fig:acc}
\end{figure}

The leading acceptance $A(K_{2\pi})$ as a function of the assumed DP mass is shown in Fig.~\ref{fig:acc}, while the ${\cal B}(K_{\mu 3})A(K_{\mu 3})$ term is negligible in the absence of a $K_{\mu 3}$ selection. However $\pi^0$ mesons produced in semileptonic decays can be used to improve the DP limits by developing a dedicated event selection requiring lower total momentum and large transverse momentum to account for the neutrino emission. The largest uncertainty on the computed ${\cal B}(\pi^0\to\gamma A')$ is the external one due to ${\cal B}(\pi^0)$ entering via $N_K$. It amounts to 3\% in relative terms and is neglected. The obtained upper limits on ${\cal B}(\pi^0\to\gamma A')$ are ${\cal O}(10^{-6})$ and do not exhibit a strong dependence on the assumed DP mass, as the negative trends in background fluctuation (Fig.~\ref{fig:observed}) and acceptance (Fig.~\ref{fig:acc}) largely cancel out.

\begin{figure}[tb]
\vspace{3mm}
\begin{flushleft}
\resizebox{0.46\textwidth}{!}{\includegraphics{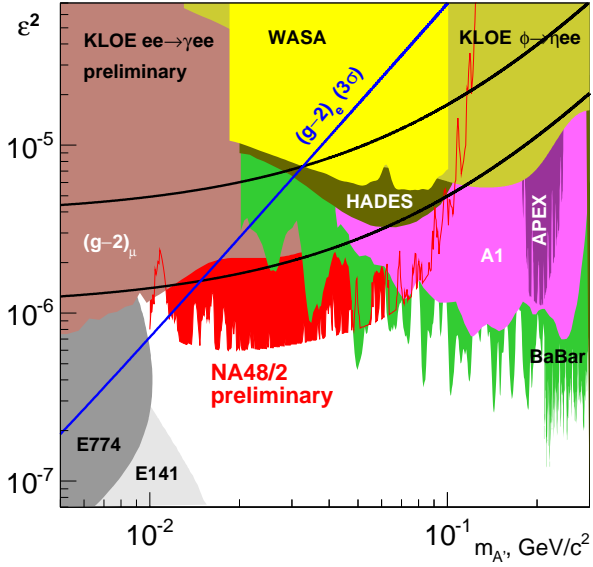}}
\end{flushleft}
\vspace{-5mm}
\caption{The NA48/2 preliminary upper limits at 90\% CL on the mixing parameter $\varepsilon^2$ versus the DP mass $m_{A'}$, compared to the other published exclusion limits from meson decay, beam dump and $e^+e^-$ collider experiments~\cite{le14}. Also shown are the band where the consistency of theoretical and experimental values of muon $g-2$ improves to $\pm 2\sigma$ or less, and the region excluded by the electron $g-2$ measurement~\cite{po09,da14}.}
\label{fig:world}
\end{figure}

Upper limits at 90\% CL on the mixing parameter $\varepsilon^2$ in each considered DP mass hypothesis are calculated from those on ${\cal B}(\pi^0\to\gamma A')$ using Eq.~(\ref{eq:br}). The resulting preliminary DP exclusion limits, along with constraints from other experiments~\cite{le14}, the band of phase space where the discrepancy between the measured and calculated muon $g-2$ values falls into the $\pm2\sigma$ range~\cite{po09,da14} due to the DP contribution, and the region excluded by the electron $g-2$ measurement, are presented in Fig.~\ref{fig:world}. The obtained upper limits on $\varepsilon^2$ represent an improvement over the existing data in the DP mass range 10--60~MeV/$c^2$, and exclude the muon $g-2$ band in the range 10--100~MeV/$c^2$. The most stringent limits ($6\times 10^{-7}$) are achieved at $m_{A'}\approx 20~{\rm MeV}/c^2$ where the acceptance for the full decay chain is the highest (reaching 2.5\%). The limits weaken at higher $m_{A'}$ due to both the kinematic suppression of the $\pi^0\to \gamma A'$ decay and the decreasing acceptance.

The assumption of prompt DP decay that is fundamental to this analysis is justified a posteriori by the obtained results: all upper limits on $\varepsilon^2 m_{A'}^2$ are above $6\times 10^{-5}~({\rm MeV}/c^2)^2$, corresponding to maximum DP mean paths in the NA48/2 reference frame below 10~cm (see Section~\ref{sec:intro}). The corresponding loss of efficiency of the trigger and event selection (both relying on 3-track vertex reconstruction) is negligible, as the typical resolution on the vertex longitudinal coordinate in the forward NA48/2 geometry is $\approx 1$~m.

\section{Summary and outlook}

The NA48/2 experiment at CERN was exposed to about $2\times 10^{11}$ $K^\pm$ decays in flight in 2003--2004. The large integrated kaon flux makes it a precision kaon by also $\pi^0$ physics facility, and the studies of the $\pi^0$ decay physics with the NA48/2 data have started. Preliminary results on dark photon search in $\pi^0$ decays are reported: no signal is observed, and the obtained upper limits on the mixing parameter $\varepsilon^2$ improve over the world data in the mass range 10--60~MeV/$c^2$. In particular, the limits at 90\% CL are $\varepsilon^2<10^{-6}$ for $12~{\rm MeV}/c^2<m_{A'}<55~{\rm MeV}/c^2$, and the strongest limits reach $\varepsilon^2=6\times 10^{-7}$ at $m_{A'} \approx 20~{\rm MeV}/c^2$. Combined with the other available data, this result rules out the DP as an explanation for the muon $(g-2)$ anomaly, assuming DP couples to quarks and decays predominantly into SM fermions.

The performed search for the prompt $A'\to e^+e^-$ decay is limited by the irreducible $\pi^0_D$ background: the obtained upper limits on $\varepsilon^2$ in the mass range 10--60~MeV/$c^2$ are about three orders of magnitude higher than the single event sensitivity. The sensitivity to $\varepsilon^2$ achievable with the employed method scales as the inverse square root of the integrated beam flux, and therefore this technique is unlikely to advance much below $\varepsilon^2=10^{-7}$ in the near future, either by improving on the NA48/2 analysis or by exploiting larger future $\pi^0$ samples (e.g. the one expected to be collected by the NA62 experiment at CERN~\cite{ruggiero}). On the other hand, a search for a long-lived (i.e. low $m_{A'}$ and low $\varepsilon^2$) DP produced in the $\pi^0$ decay from high momentum kaon decay in flight using the {\it displaced vertex} method would be limited by the $\pi^0_D$ background to a lesser extent, and its sensitivity is worth investigating.

%
%

\end{document}